\documentclass[conference]{IEEEtran}
\IEEEoverridecommandlockouts
\usepackage{cite}
\usepackage{amsmath,amssymb,amsfonts}
\usepackage{algorithmic}
\usepackage{graphicx}
\usepackage{textcomp}
\usepackage{xcolor}
\usepackage{cleveref}
\usepackage{lipsum} 
\usepackage{booktabs}
\usepackage{stfloats}
\usepackage{multirow}
\usepackage{tabularx}
\usepackage{subcaption}
\usepackage{caption}
\usepackage{mathtools}
\usepackage[acronyms]{glossaries}
\makeglossaries

\newcommand{\newdefineabbreviation}[4]
{
    \newglossaryentry{#1_glossary}
    {
        text={#2},
        long={#3},
        name={\glsentrylong{#1_glossary} (\glsentrytext{#1_glossary})},
        description={#4}
    }

    \newglossaryentry{#1}
    {
        type=\acronymtype,
        name={\glsentrytext{#1_glossary}}, 
        description={\glsentrylong{#1_glossary}}, 
        first={\glsentryname{#1_glossary}\glsadd{#1_glossary}},
    }
}

\newdefineabbreviation{uav}{UAV}{Unmanned Aerial Vehicle}{unmanned aerial vehicle}
\newdefineabbreviation{fso}{FSO}{Free-Space Optical}{free-space optical}
\newdefineabbreviation{fsm}{FSM}{\emph{Fast Steering Mirror}}{fast steering mirror}
\newdefineabbreviation{fov}{FOV}{Field of View}{field of view}
\newdefineabbreviation{rf}{RF}{Radio Frequency}{radio frequency}
\newdefineabbreviation{e2e}{E2E}{End-to-End}{end-to-end}
\newdefineabbreviation{pdf}{PDF}{Probability Density Function}{Probability Density Function}
\newdefineabbreviation{aci}{ACI}{Adaptive Channel and switch-aware Index}{adaptive channel- and switch-aware index}
\newdefineabbreviation{mw}{MW}{Max-Weight}{Max-Weight}
\def\BibTeX{{\rm B\kern-.05em{\sc i\kern-.025em b}\kern-.08em
    T\kern-.1667em\lower.7ex\hbox{E}\kern-.125emX}}
\begin{document}

\title{Dynamic Server Allocation Under Stochastic Switchover on Time-Varying Links\\}

\author{\IEEEauthorblockN{Hossein Mohammadalizadeh}
\IEEEauthorblockA{\textit{Hasso-Plattner-Institute} \\
\textit{Internet Technologies and Softwarization} \\
Potsdam, Germany \\
ho.mohammadalizadeh@hpi.de}
\and
\IEEEauthorblockN{Holger Karl}
\IEEEauthorblockA{\textit{Hasso-Plattner-Institute} \\
\textit{Internet Technologies and Softwarization} \\
Potsdam, Germany \\
holger.karl@hpi.de}
}

\maketitle

\begin{abstract}
Dynamic resource allocation to parallel queues is a cornerstone of network scheduling, yet classical solutions often fail when accounting for the overhead of switching delays to queues with superior link conditions. In particular, system performance is further degraded when switching delays are stochastic and inhomogeneous. In this domain, the myopic, Max-Weight policy struggles, as it is agnostic to switching delays. This paper introduces \gls{aci}, a non-myopic, frame-based scheduling framework that directly amortizes these switching delays. We first use a Lyapunov drift analysis to prove that backlog-driven \gls{aci} is throughput-optimal with respect to a scaled capacity region; then validate \gls{aci}'s effectiveness on multi-UAV networks with an \gls{fso} backhaul. Finally, we demonstrate how adapting its core urgency metric provides the flexibility to navigate the throughput-latency trade-off.
\end{abstract}

\section{Introduction}
\label{sec:intro}

\begingroup
  \renewcommand\thefootnote{}%
  \footnotetext{The authors acknowledge the financial support by the Federal Ministry
of Education and Research of Germany in the project “Open6GHub” (grant
number: 16KISK011)}%
  \addtocounter{footnote}{-1}%
\endgroup

A broad class of communication systems can be cast as dynamic resource allocation to parallel queues subject to \emph{time-varying link conditions} and \emph{inhomogeneous switching delays} \cite{212277}. The central challenge is to design a scheduling policy that is both channel-aware (accounting for link conditions) and switch-aware (accounting for switching delays) to ensure system stability and maximize throughput. Existing paradigms, however, often idealize away these difficulties, leaving a gap between theory and the reality of mobile or directional platforms. This work addresses that gap by formalizing a single-server, parallel-queue model that incorporates random, pair-dependent switching delays and time-varying channels. Within this setting, our goal is to develop scheduling policies that balance instantaneous rate against switching delays. To ground this abstract problem, we introduce a practical example from \gls{uav}-assisted \gls{fso} communication that exhibits these exact dynamics.

\begin{figure}[h!]
    \centering
    \includegraphics[keepaspectratio, width=\columnwidth, height=5cm]{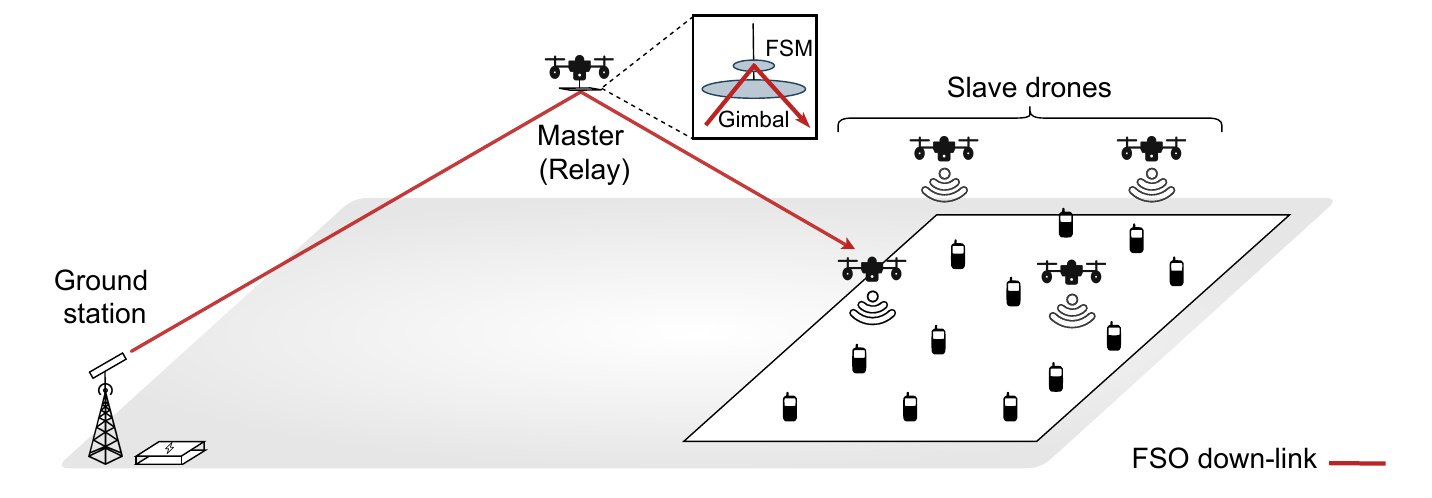}
    \caption{System model}
    \label{fig:big-picture}
\end{figure}

Consider a two-hop \gls{uav}-assisted wireless network with an \gls{fso} backhaul, comprising a ground station, one \emph{master} \gls{uav}, and multiple \emph{slave} \glspl{uav} (see \cref{fig:big-picture}). In this system, the ground station points an optical beam to the master drone, which acts as a relay, actively steering the beam to a single chosen slave per time-slot. This physical setup directly gives rise to the two central challenges of our model.

First, the link quality is inconsistent. The combination of atmospheric dynamics, platform mobility, and potential blockages—compounded by the mechanical limitations of steering and the slaves' finite receiver \gls{fov}—induces the time-varying channel conditions. Second, switching between slaves is not instantaneous. The handover process requires coarse steering with a \emph{gimbal mirror} and fine alignment with a \gls{fsm}, resulting in a delay whose duration depends on the slaves' angular separation \cite{8452986}. This mechanical process is the physical source of the stochastic and inhomogeneous switching delays. 
This example faithfully manifests the structural features of the abstract problem: time-varyting connectivity (stemming from alignment, \gls{fov}, and blockage) and inhomogeneous switching (stemming from steering and acquisition). Additionally, the constraint of a single active link makes the choice of scheduling policy important, as it directly dictates network stability and throughput.

Building on this motivation, our contributions are as follows: We (i) provide a concrete system-to-model instantiation that captures the twin challenges of dynamically allocating resources to parallel queues with time-varying connectivity and inhomogeneous switching; (ii) characterize the \gls{fso} channel variability and switching statistics induced by actuator dynamics, alignment, and \gls{fov} constraints; (iii) we propose and analyze scheduling policies that adapt jointly to channel and switching dynamics and prove throughput optimality—i.e., stabilization of any arrival rate vector strictly within the capacity region under standard ergodicity and finite-moment assumptions; and (iv) validate our approach with simulations.

\section{Related Work}
\label{sec:rw}
The problem of dynamic server allocation to parallel queues is foundationally addressed by the \gls{mw} algorithm, which is proven to be throughput-optimal when server switching is instantaneous and cost-free \cite{182479}. However, in many practical systems, switching server incurs significant time delays. The introduction of such switchover times fundamentally alters the problem, shrinking the stability region and rendering the myopic \gls{mw} policy sub-optimal, and potentially unstable \cite{6236167}. Subsequent research has developed switch-overhead-aware policies to address this. One major approach is to amortize the switching cost by committing to a service decision for an extended period. This includes frame-based policies like the FBDC algorithm, which regains asymptotic optimality for fixed switchover delays by using long service dwells \cite{6236167}, and two-timescale policies that separate slow base-station activation from fast user scheduling in wireless networks \cite{8056983, 8737401}. A second approach uses hysteresis to suppress inefficient flapping, where the scheduler waits for a state-dependent backlog threshold to be crossed before initiating a switch \cite{Chan2016-fb}. These seminal works provide robust solutions but predominantly model switching as a deterministic, uniform delay or an abstract cost penalty therefore they do not fit the scenario in \cref{sec:intro}, where switching times are state-dependent and stochastic. Our methodology addresses this gap by directly integrating these state-dependent, stochastic switching times into a Lyapunov-based scheduling policy, thereby creating a model that is better aligned with the realities of queuing systems.

\section{System Model}
\label{sec:sysmodel}
\subsection{Queueing Formulation}
\label{sec:netarch}

We study an \textbf{M/G/1} queuing system with one server and $|\mathcal{N}|$ parallel queues, each
with Markovian arrivals and general service times. Time is slotted into $T$ intervals of length
$\Delta t$, indexed by $t$; in each slot the server serves at most one queue. Switching from queue
$i$ to $j$ incurs a switching of $\tau_{ij}\in\mathbb{Z}^+$ slots. Let $Q_i(t)$ denote the backlog of queue $i$ at the start of slot $t$, with Poisson arrivals
$A_i(t)$ of rate $\lambda_i$ and infinite buffers. The server--queue assignment is given by
$a_i(t)\in\{0,1\}$ ($1=$ scheduled), subject to $\sum_{i\in\mathcal{N}} a_i(t)\le 1$ for all $t$.
Server unavailability (due to switching or other outages) is represented by $b(t)\in\{0,1\}$
($b(t)=1$ means unavailable), and service eligibility is $e_i(t)=a_i(t)\,(1-b(t))$. The instantaneous
channel rate of queue $i$ in slot $t$ is $R_i(t)$, yielding effective per-slot capacity
$\mu_i(t)=\min\{\bar{\mu},\,R_i(t)\}$, where $\bar{\mu}$ is the system throughput cap. Departures and
backlog evolve as:
\begin{align}
  D_i(t) &= \big(Q_i(t)\wedge \mu_i(t)\big)\,e_i(t), \label{eq:departures}\\
  Q_i(t+1) &= \big(Q_i(t)-D_i(t)\big)^+ + A_i(t),\quad Q_i(0)=0 \label{eq:queue-evolve}
\end{align}
with $x\wedge y=\min\{x,y\}$ and $(x)^+=\max\{x,0\}$.
We model switching delays, $\tau_{ij}(t)$, as \emph{dependent random variables} because, in practice, the time required to switch is not an isolated event. It's influenced by the specific source and destination queues as well as recent switching activity. This approach is superior to modeling the delays as \emph{independent random variables} because it correctly captures how a scheduling policy's choices impact future switching costs. While simpler, the independent model ignores this.

A scheduling policy $\pi$ selects the next target queue at slot boundaries from the
system state:
\begin{equation}
  \mathcal{S}(t) \;=\;
  \big[\, [Q_i(t)]_{i\in\mathcal{N}},\; [R_i(t)]_{i\in\mathcal{N}},\; [\tau_{ij}(t)]_{i\ne j} \,\big]
\end{equation}
which includes current backlogs, instantaneous channel rates, and pairwise
switching delays. The policy induces per-slot decisions $(a_i(t),b(t))$; when
switching from $i$ to $j$, it sets $b(t)=1$ for $\tau_{ij}(t)$ consecutive slots
(so $e_i(t)=0$ for all $i$), after which scheduling resumes with $a_j(t)=1$.

Under a fixed policy with long-run switching frequencies $p_{ij}$ (moves $i\!\to\! j$)
and average visit lengths $v_i$, the fraction of slots lost to switching/outages is:
\begin{equation}
  \phi_{\text{sw}} \;=\;
  \frac{\sum_{i\ne j} p_{ij}\,\bar{\tau}_{ij}}{\sum_{k=1}^N v_k \;+\; \sum_{i\ne j} p_{ij}\,\bar{\tau}_{ij}}
  \label{eq:phi-sw}
\end{equation}
so the usable fraction of time is $1-\phi_{\text{sw}}$. Each queue $i$ delivers
average per-slot service $r_i=\mathbb{E}[\mu_i(t)]$ when connected and is scheduled
in a fraction $\alpha_i$ of the usable slots; its effective allocation over all slots is:
\begin{align}
  \varphi_i &= (1-\phi_{\text{sw}})\,\alpha_i \label{eq:varphi-def}\\
  \varphi_i &\ge 0, \qquad \sum_{i=1}^N \varphi_i \le 1-\phi_{\text{sw}}
  \label{eq:varphi-cons}
\end{align}
Feasibility of $\Lambda=(\lambda_1,\dots,\lambda_N)$ requires:
\begin{equation}
  \lambda_i < \varphi_i\,r_i, \quad \forall i
  \label{eq:per-queue-feas}
\end{equation}
which yields the compact inner bound:
\begin{equation} 
    \mathcal{C} \;\supseteq\; \Big\{\,\Lambda\in\mathbb{R}_+^N \,:\, \sum_{i=1}^N \frac{\lambda_i}{r_i} \;<\; 1-\phi_{\text{sw}} \,\Big\} 
\end{equation}
The visit lengths $v_i$ are determined by the load and the chosen scheduling policy. 
With the preceding queueing formulation and switching-time models, a beam-steered FSO backhaul for a
multi-UAV network aligns naturally: per-slot link rates $R_i(t)$ are supplied by the FSO link, while pairwise switching times $\tau_{ij}(t)$ capture beam steering, acquisition, and
handover. In the next section we specify the FSO layer that generates $R_i(t)$ and
$\tau_{ij}(t)$—including the rate–SNR mapping, and the
(time-varying, heterogeneous) switching delays—and couple it to the queuing system above.

\subsection{Network Architecture \& \gls{fso} Channel Characteristics}
\label{subsec:fso-channel}

The network in \cref{fig:big-picture} comprises a stationary ground station, a fixed master UAV acting as a reflective relay, and $|\mathcal{N}|$ slave UAVs arranged in a large hexagonal pattern. Each slave loiters on a short circular path around its formation point to emulate continuous motion. The end-to-end (E2E) optical gain from ground to slave $i$ in slot $t$ is a two-hop product through the master, scaled by mirror reflectivity $\rho$:
\begin{equation}
H_i(t)=\rho\,H_1(t)\,H_{2,i}(t)
\label{eq:e2e_gain}
\end{equation}
Each hop $\kappa\in\{1,2\}$ factors into path loss $h_{\ell}$, atmospheric turbulence $h_a$, geometric coupling $h_g$, and pointing loss $h_p$ \cite{8478112}:
\begin{equation}
H_\kappa(t)=h_{\ell,\kappa}(t)\,h_{a,\kappa}(t)\,h_{g,\kappa}\,h_{p,\kappa}(t).
\label{hop-gains}
\end{equation}

Over a propagation range $Z$, large-scale attenuation is $h_{\ell}=\exp(-\xi Z)$ with extinction coefficient $\xi$. Weak-fluctuation scintillation is modeled as $h_a(t)=\exp(X_t)$ with $X_t\sim\mathcal{N}(\mu,\sigma^2)$, $\mu=-2V$, $\sigma^2=4V$, which yields $\mathbb{E}[h_a]=1$ and controls fading severity via the log-amplitude variance $V$. The geometric coupling under perfect alignment is:
\begin{align}
h_g &= A_0=\bigl[\mathrm{erf}(\nu)\bigr]^2
& \nu&=\frac{\sqrt{\pi}\,a}{\sqrt{2}\,w_z}
\label{eq:geom_A0}
\end{align}
and the pointing loss for instantaneous radial error $r(t)$ is:
\begin{align}
h_p(t)&=\exp\!\left(-\frac{2\,r(t)^2}{w_{\mathrm{eq}}^{\,2}}\right)
& w_{\mathrm{eq}}^{\,2}&=w_z^{2}\sqrt{\frac{\pi\,\mathrm{erf}(\nu)}{2\nu\,e^{-\nu^{2}}}}
\label{eq:pointing_weq}
\end{align}
Here, $a$ is the receiving aperture radius and $w_z$ is the gaussian beam radius and $\nu$ represents the ratio of the receiver's aperture size to the size of the incoming \gls{fso} beam.
Hop~1 (ground$\to$master) is static, with Rayleigh radial error $r_1(t)$ and per-axis jitter variance $\sigma_{pg}^2=\sigma_{p1}^2+\sigma_g^2$ from the ground transmitter and the master platform. Hop~2 (master$\to$slave) is dynamic; the total error combines lateral displacement ($\delta_{\text{lat}}(t)$) and angular deviation ($\theta_{\text{ang}}(t)$) scaled by the instantaneous range $Z_{2,i}(t)$:
\begin{align}
r_{2,i}(t) &= \delta_{\mathrm{lat}}(t)+Z_{2,i}(t)\,\theta_{\mathrm{ang}}(t)\\
\mathrm{Cov}\!\left[\delta_{\mathrm{lat}}(t)\right] &= (\sigma_{p1}^2+\sigma_{p2}^2)\,I_2\\
\mathrm{Cov}\!\left[\theta_{\mathrm{ang}}(t)\right] &= \bigl(4\sigma_{\theta,m}^2+\sigma_{\theta,2}^2+\sigma_{\mathrm{turb}}^2\bigr)\,I_2
\label{eq:hop2_error_vector}
\end{align}
where mirror jitter at the master enters doubled in angle (factor $2$; variance factor $4$). A hard \gls{fov} gate applies: the hop gain is zero if $\|\mathbf{r}_{2,i}(t)\|>Z_{2,i}(t)\theta_{\mathrm{FOV}}$. Table~\ref{tab:channel_params_condensed} summarizes hop-specific quantities.

\begin{table}[h]
\centering
\renewcommand{\arraystretch}{1.25}
\caption{Hop-specific parameters}
\label{tab:channel_params_condensed}
\begin{tabular}{l l l}
\toprule
\textbf{Parameter} & \textbf{Hop 1 (Static)} & \textbf{Hop 2 (Dynamic)}\\
\midrule
Link distance & fixed $Z_1$ & time-varying $Z_{2,i}(t)$\\
Optics $(a,w_z)$ & $(a_m,\,w_{z,1})$ & $(a_r,\,w_{z,2})$\\
Per-axis jitter & $\sigma_{pg}^2=\sigma_{p1}^2+\sigma_g^2$ 
& $\sigma_{\mathrm{lat}}^2=\sigma_{p1}^2+\sigma_{p2}^2$\\
Angular jitter & — & $\sigma_{\mathrm{ang}}^2=(2\sigma_{\theta,m})^2+\sigma_{\theta,2}^2$\\
FOV gate & none & $\|\mathbf{r}_{2,i}(t)\|\le Z_{2,i}(t)\theta_{\mathrm{FOV}}$\\
\bottomrule
\end{tabular}
\end{table}

The distribution of the E2E gain $H_i(t)$ lacks a closed form due to the product structure and the nonlinear FOV truncation. It is therefore estimated via Monte Carlo simulation of the constituent terms, and the outage probability follows as:
\begin{equation}
P_{\mathrm{out}}(h_{\mathrm{th}})=\Pr[H_i(t)<h_{\mathrm{th}}]=\int_{0}^{h_{\mathrm{th}}} f_{H_i}(h)\,dh,
\label{eq:p-outage}
\end{equation}
with the gain threshold tied to the minimum decoding SNR by:
\begin{equation}
h_{\mathrm{th}}=\frac{\sqrt{\mathrm{SNR}_{\min}}\,\sigma_n}{R \cdot P_t}
\label{eq:h_thresh}
\end{equation}
where $R$ is the detector responsivity, $P_t$ the transmit power, and $\sigma_n$ the receiver noise standard deviation. The resulting instantaneous data rate for slave $i$ is:
\begin{equation}
R_i(t)=\eta\,B_e\log_2\!\left(1+\frac{\mathrm{SNR}_i(t)}{\Gamma}\right)
\label{eq:rate_final}
\end{equation}
with efficiency $\eta$, electrical bandwidth $B_e$, and SNR gap $\Gamma$ relative to the Shannon limit.
The channel is time-selective, with its coherence time governed by atmospheric turbulence and platform jitter. The atmospheric coherence time, $t_0$, is given by \cite{Tokovinin2008-lp}:
\begin{equation}
\label{eq:greenwood}
t_0 \approx \left[ 2.91 k^2 \int_{\text{path}} C_n^2(h) v(h)^{5/3} dh \right]^{-3/5}
\end{equation}
where $k$ is the optical wave number, and $C_n^2(h)$ and $v(h)$ are the refractive index and wind speed profiles. As the coherence time of atmospheric effects ($t_0 \approx 10$\,ms) is much shorter than that of mechanical platform jitter, turbulence is the dominant factor driving rapid channel fluctuations. This justifies a \textbf{block-fading model}, where the channel gain is assumed to be constant over a slot duration $\Delta t$ on the order of $t_0$.

\subsection{Stochastic Switching Time Model}
\label{sec:switchmodel}
Switching the optical beam from a serving slave $i$ to a target slave $j$ is not instantaneous. The total switching time $\mathcal{T}_{ij}(t)$ is a random variable to account for mechanical, electronic, and probabilistic effects. In our example scenario, this time accounts for the sum of the time to re-point the gimbal and \gls{fsm} and the total time for all link acquisition attempts:
\begin{equation}
\label{eq:switching_time}
\mathcal{T}_{ij}(t) = T_{\text{slew}}(\theta_{ij}) + K \cdot T_{\text{acq}}, \quad \tau_{ij}(t) = \left\lceil\frac{{\mathcal{T}}_{ij}(t)}{\Delta t}\right\rceil
\end{equation}
The first term, $T_{\text{slew}}$, is the deterministic mechanical slew time. It is governed by a trapezoidal S-curve velocity profile based on the angular separation $\theta_{ij}$ and the gimbal's dynamic limits on maximum velocity ($v_{\text{max}}$), acceleration ($a_{\text{max}}$), and jerk ($j_{\text{max}}$):
\begin{equation}
\label{eq:slew_time}
T_{\text{slew}}(\theta_{ij}) = \frac{\theta_{ij}}{v_{\text{max}}} + \frac{v_{\text{max}}}{a_{\text{max}}} + \frac{4a_{\text{max}}}{j_{\text{max}}}
\end{equation}
The second term in \cref{eq:switching_time} contains the fixed overhead
for a single acquisition attempt, $T_{\text{acq}} = t_{\text{fsm}} + t_{\text{pilot}}$, which includes fine-pointing FSM settling and pilot signal exchange. The total time to establish a link is the number of attempts, $K$, multiplied by the fixed duration of a single attempt, $T_{\text{acq}}$. We model $K$ as a \textbf{geometrically distributed} random variable where the success probability, $p_i(t)$, is dynamic, depending on factors like link distance and pointing accuracy. This model accounts for both \textbf{stochastic failures}, where multiple attempts ($K>1$) increase the delay, and \textbf{geometric rejections}, where a link is physically impossible ($p_i(t)=0$). Since this underlying success probability changes over time, the resulting sequence of switching delays is \textbf{temporally correlated}.
\section{Scheduling Approaches}
\label{sec:approaches}
We begin by recalling the classical \gls{mw} policy. \gls{mw} is provably throughput-optimal when switching costs are absent\cite{182479}; we therefore treat it as a comparison case that captures the best achievable performance in an idealized, no-cost setting. Real systems, however, face inhomogeneous switching delays that \gls{mw} ignores. To address this gap, we introduce our \textbf{Adaptive Channel and switch-aware Index (ACI)} scheduler, which is explicitly designed for settings with stochastic switching delays. We prove that \gls{aci} remains throughput-optimal and then develop variations to optimize additional performance objectives under these realistic costs.

\subsection{The Max-Weight Algorithm: A Theoretical Benchmark}
\label{subsec:max_weight}

The classical \gls{mw} policy is throughput-optimal for parallel queues \emph{when switching costs are zero}. In our discrete-time model (Sec.~\ref{sec:netarch}), a single server serves queue $i\in\mathcal{N}$ with backlog $Q_i(t)$ and instantaneous service rates $\mu_i(t)$ when connected. At each slot $t$, \gls{mw} selects:
\begin{equation}
\label{eq:max_weight_rule}
i^*(t) \in \underset{i \in \mathcal{N}}{\mathrm{argmax}}\; \big[Q_i(t) \cdot \mu_i(t) \cdot \Delta t\big]
\end{equation}
This balances backlog and rate and, in the zero-cost case, stabilizes any arrival vector $\Lambda$ within the capacity region $\mathcal{C}$. Because \gls{mw} ignores switching delays, applying it with heterogeneous switching delays can trigger frequent handovers and waste service time. This motivates our \gls{aci} scheduler developed next.

\subsection{The ACI Scheduling Framework}
\label{subsec:aci_framework}
ACI departs from MW in two ways: (i) it accounts for switching delays in the decision rule, and (ii) it commits to frames of length $L$ instead of single-slot decisions. Frames both amortize the switching delay over $L$ slots and let non-served queues accumulate arrivals (increasing their weights), and reduces oscillations between queues. A switch occurs only when its expected service gain exceeds the switching delay. At each decision point \gls{aci} selects queue $i^*$ that maximizes a generalized score:
\begin{align}
i^*(t) &= \underset{i \in N}{\mathrm{argmax}}\;
          \Big[Q_i(t)\cdot\bar{\mu}_{i|j}(t)\cdot f_{ij}(t)\Big] \label{eq:aci_rule}
\end{align}
The score is a product of three components. $Q_i(t)$ is the backlog of slave $i$ at time $t$. $\bar{\mu}_{i|j}(t)$ is the amortized goodput. Ideally, this would be based on the true expected goodput in bits over the frame, $B_i(t;L)$, which requires integrating over future channel rates:
\begin{equation}
\label{eq:total_bits_expected}
B_i(t;L) \triangleq \sum_{\ell=0}^{L-1} R_i(t+\ell \Delta t) \cdot (\Delta t - t_p)^+
\end{equation}
Here, $\Delta t$ is the slot duration and $t_p \in [0,\Delta t)$ is the per-slot processing overhead), so only $(\Delta t - t_p)^+$ seconds per slot are usable.
Since future rates $R_i(t+\ell \Delta t)$ are unknown at the decision time $t$, a practical scheduler must approximate this expectation. We use the current rate $R_i(t)$ as a forecast for the entire dwell:
\begin{equation}
\label{eq:total_bits_approx}
\hat{B}_i(t;L) \approx L \cdot R_i(t) \cdot (\Delta t - t_p)^+
\end{equation}
To add robustness, \gls{aci} supports \emph{early halt}. Service on the current queue may end before the frame completes when any of the following holds: (i) the queue drains; (ii) the channel is in persistent outage; (iii) the realized rate falls consistently below the forecast used in \cref{eq:total_bits_approx}; or (iv) another queue’s score in \cref{eq:aci_rule} grows sufficiently larger than the current one. Because the score is continuously recalculated, the scheduler can preempt frames when these conditions hold. The amortized goodput is then the initially-estimated total bits, divided by the total time investment (switching time $\tau_{ij}(t)$ and the planned dwell $L \cdot \Delta t$):
\begin{equation}
\label{eq:amortized_goodput_final}
\bar{\mu}_{i|j}(t) = \frac{\hat{B}_i(t;L)}{\tau_{ij}(t) + L \cdot \Delta t}
\end{equation}
Finally, ACI provides explicit control via a \textbf{Switching Modulator}, $f_{ij}(t)$, which is central to the scheduling decision and is defined as:
\begin{equation}
f_{ij}(t) \triangleq \frac{1 + \gamma \cdot \chi_{ij}(t)}{1 + \beta \cdot \tau_{ij}(t)}
\end{equation}
This term combines a bonus for \emph{transition affinity}, $\chi_{ij}(t) \in [0,1]$, with a penalty for the switching duration in slots, $\tau_{ij}(t)$. Transition affinity is a normalized metric that quantifies the efficiency of switching from queue $i$ to $j$, based on the system's physical or logical structure. For instance, in a network of mobile UAVs where proximity reduces switching time, affinity is physical adjacency. Alternatively, in a system where queues rely on shared software modules, affinity is the degree of resource overlap. The trade-off between these factors is controlled by $\gamma, \beta \ge 0$: increasing $\gamma$ prioritizes high-affinity switches, while increasing $\beta$ more heavily penalizes the raw duration to favor faster switchings. For our subsequent optimality proof, we establish its bounds. Given a maximum possible switching duration $\tau_{\max}$, the modulator is bounded, where $f_{\max} = 1 + \gamma$ and $f_{\min} = \tfrac{1}{1 + \beta \tau_{\max}}$.

\subsection{Throughput Optimality of the ACI Scheduler}
\label{subsec:throughput_optimality}
We prove that ACI is throughput-optimal by showing that the network queues are stable for any arrival rate vector within a scaled version of the capacity region. The proof uses Lyapunov criterion, which requires showing that the system state is pushed towards a smaller-congestion region when it becomes large. To formalize this, we first establish a bound on the one-slot drift of a quadratic Lyapunov function, defined as a measure of total network congestion:
\begin{align}
    L(Q(t)) \triangleq \frac{1}{2} \sum_{i=1}^{N} Q_i^2(t)
    \label{eq:lyapunov}
\end{align}
The one-slot drift, $\Delta(t)$, is its expected change over a single time slot, conditioned on the current state:
\begin{equation}
    \Delta(t) \triangleq \mathbb{E}\!\left[L(Q(t+1)) - L(Q(t)) \,\big|\, Q(t)\right]
    \label{eq:drift_def}
\end{equation}
We derive a bound on this drift by analyzing the change in a single queue's squared backlog. For conciseness, we drop the time and queue indices, letting $Q_i(t+1)=Q^+$, $Q_i(t)=Q$, $A_{i}(t)=A$, $D_{i}(t)=D$, and the remaining backlog be $U=[Q-D]^+$. The derivation begins from the queueing dynamic $Q^+=U+A$:
\begin{align}
    \label{eq:drift_expand_start}
    (Q^+)^2 = (U+A)^2 = U^2 + A^2 + 2AU
\end{align}
We can upper-bound this expression by applying two inequalities. First, since the remaining backlog $U = [Q-D]^+$ cannot exceed the current queue size, we have $U \le Q$. Second, because the square of the positive part of a number is no larger than the square of the number itself, it implies $U^2 \le (Q-D)^2$. Applying these bounds yields:
\begin{align}
    \label{eq:drift_expand_end}
    (Q^+)^2 &\le (Q-D)^2 + A^2 + 2AQ \nonumber \\
    &= Q^2 - 2QD + D^2 + A^2 + 2AQ
\end{align}
Rearranging this expression gives the bound on the per-queue change:
\begin{equation}
    \label{eq:per_queue_bound}
    (Q^+)^2 - Q^2 \le A^2 + D^2 + 2Q(A - D)
\end{equation}
Substituting this result back into the definition of the drift from \cref{eq:drift_def} and restoring the indices yields the final bound:
\begin{align}
    \label{eq:drift_bound}
    \Delta(t) &\le \mathbb{E}\bigg[ \frac{1}{2}\sum_{i=1}^N \Big(A_i(t)^2 + D_i(t)^2\Big) \nonumber\\
    &{}+ \sum_{i=1}^N Q_i(t)\Big(A_i(t) - D_i(t)\Big) \,\Big|\, Q(t) \bigg]
\end{align}
To analyze the system over service frames, we extend the one-slot drift bound. We hereafter let $\sum_t$ denote the sum over all slots in frame $k$, i.e., $\sum_{t=t_k}^{t_{k+1}-1}$. Summing \cref{eq:drift_bound} over a frame of length $F_k = t_{k+1}-t_k$ gives:
\begin{equation}
\begin{aligned}
    L(Q(t_{k+1})) - L(Q(t_k)) &\le \frac{1}{2}\sum_t\sum_{i=1}^N \!\big(A_i(t)^2+D_i(t)^2\big) \\
    &\quad+\sum_t\sum_{i=1}^N \! Q_i(t)\big(A_i(t)-D_i(t)\big)
\end{aligned}
\label{eq:frame_sum}
\end{equation}
The main challenge is to bound the cross-term involving $Q_i(t)$, as it evolves within the frame. To do this, we first linearize the queueing dynamic.

Let the net input be $\delta_i(t) \triangleq A_i(t)-D_i(t)$, and define a non-negative slack variable $\eta_i(t) \triangleq (D_i(t)-Q_i(t))^{+}$ to capture the effect of the non-negativity constraint. This rewrites the queueing dynamic in a linear form:
\begin{equation}
    Q_i(t+1) = Q_i(t) + \delta_i(t) + \eta_i(t)
    \label{eq:queue-linear}
\end{equation}
Crucially, the slack is bounded by the service, $\eta_i(t) \le D_i(t)$. By unrolling the recursion, we express $Q_i(t)$ in terms of the queue state at the start of the frame, $Q_i(t_k)$:
\begin{equation}
    Q_i(t) = Q_i(t_k) + \sum_{u=t_k}^{t-1}\big(\delta_i(u)+\eta_i(u)\big)
    \label{eq:Q-expand}
\end{equation}
We substitute the expansion of $Q_i(t)$ from \cref{eq:Q-expand} into the cross-term. Distributing the $\delta_i(t)$ term and factoring out the constant $Q_i(t_k)$ separates the expression into three distinct components: a frame-start term and two remainder terms, which we define as $R_i^{\delta\delta}$ and $R_i^{\delta\eta}$:
\begin{align}
    \sum_{t} Q_i(t)\delta_i(t) &= Q_i(t_k)\sum_{t}\delta_i(t) + \underbrace{\sum_{t}\left(\sum_{u<t}\delta_i(u)\right)\delta_i(t)}_{\triangleq R_i^{\delta\delta}} \nonumber \\
    &\qquad + \underbrace{\sum_{t}\left(\sum_{u<t}\eta_i(u)\right)\delta_i(t)}_{\triangleq R_i^{\delta\eta}}
    \label{eq:cross-decompose-labeled}
\end{align}

The remainder terms can now be bounded. We analyze each term separately. The term $R_i^{\delta\delta}$ is a sum over ordered pairs. It can be expressed exactly using the identity relating it to the square of the sum over the frame:
\begin{equation}
    R_i^{\delta\delta} = \frac{1}{2}\left[\left(\sum_{t}\delta_i(t)\right)^2-\sum_{t}\delta_i(t)^2\right]
    \label{eq:Rdd-identity-app}
\end{equation}
Regarding the second reminder, we can bound the first term on the right-hand side using the Cauchy-Schwarz inequality, which states $(\sum_t x_t)^2 \le F_k \sum_t x_t^2$. This gives:
\begin{align}
    R_i^{\delta\delta} &\le \frac{1}{2}\left[F_k\sum_{t}\delta_i(t)^2 - \sum_{t}\delta_i(t)^2\right] \nonumber \\
    &= \frac{F_k-1}{2}\sum_{t}\delta_i(t)^2
    \label{eq:Rdd-bound-pre}
\end{align}
Finally, using the inequality $(a-b)^2 \le 2(a^2+b^2)$, we have $\delta_i(t)^2 = (A_i(t)-D_i(t))^2 \le 2(A_i(t)^2+D_i(t)^2)$. Substituting this yields a bound in terms of second moments:
\begin{equation}
    R_i^{\delta\delta} \le (F_k-1)\sum_{t}\big(A_i(t)^2+D_i(t)^2\big)
    \label{eq:Rdd-bound-final}
\end{equation}
We start by taking the absolute value of $R_i^{\delta\eta}$ and applying the triangle inequality. Since $\eta_i(u) \ge 0$, the cumulative sum is non-negative and bounded by the total sum over the frame:
\begin{align}
    |R_i^{\delta\eta}| &= \left|\sum_{t}\left(\sum_{u<t}\eta_i(u)\right)\delta_i(t)\right| \le \sum_{t}\left(\sum_{u<t}\eta_i(u)\right)|\delta_i(t)| \nonumber \\
    &\le \left(\sum_{u}\eta_i(u)\right) \left(\sum_{t}|\delta_i(t)|\right)
    \label{eq:Rde-separate}
\end{align}
Next, we replace the terms with their upper bounds: $\eta_i(u) \le D_i(u)$ and $|\delta_i(t)| = |A_i(t)-D_i(t)| \le A_i(t)+D_i(t)$. This gives:
\begin{equation}
    |R_i^{\delta\eta}| \le \left(\sum_{u}D_i(u)\right) \left(\sum_{t}(A_i(t)+D_i(t))\right)
    \label{eq:Rde-replace}
\end{equation}
Applying the Cauchy-Schwarz inequality to each sum yields:
\begin{align}
    |R_i^{\delta\eta}| &\le \left(\sqrt{F_k}\Big(\sum_u D_i(u)^2\Big)^{\frac{1}{2}}\right) \nonumber \\
    &\qquad \times \left(\sqrt{F_k}\Big(\sum_t (A_i(t)+D_i(t))^2\Big)^{\frac{1}{2}}\right) \nonumber \\
    &\le \sqrt{2} F_k \Big(\sum_u D_i(u)^2\Big)^{\frac{1}{2}} \Big(\sum_t (A_i(t)^2+D_i(t)^2)\Big)^{\frac{1}{2}}
    \label{eq:Rde-cs}
\end{align}
where the second step again uses $(a+b)^2 \le 2(a^2+b^2)$. Finally, using the inequality $xy \le \frac{1}{2}(x^2+y^2)$ on the two square-root terms, we arrive at a linear bound:
\begin{equation}
    |R_i^{\delta\eta}| \le c_1 F_k \sum_{t}\big(A_i(t)^2+D_i(t)^2\big)
    \label{eq:Rde-bound-final}
\end{equation}
for some constant $c_1 > 0$. Combining the bounds from \cref{eq:Rdd-bound-final} and \cref{eq:Rde-bound-final} confirms that the sum of the remainders is bounded as stated.
The remainder terms can now be bounded. They are found to be proportional to the frame length and second moments of arrivals and departures:
\begin{equation}
    |R_i^{\delta\delta}| + |R_i^{\delta\eta}| \le C F_k \sum_{t}\big(A_i(t)^2+D_i(t)^2\big)
    \label{eq:remainders-bound}
\end{equation}
We now assemble the final bound by substituting our results back into the frame sum \cref{eq:frame_sum}. First, we replace the cross-term $\sum_t Q_i(t)\delta_i(t)$ with its decomposition from \cref{eq:cross-decompose-labeled}.
\begin{align}
    L(Q(t_{k+1})) - L(Q(t_k)) \le \frac{1}{2}\sum_t\sum_{i=1}^N \!\big(A_i(t)^2+D_i(t)^2\big) \nonumber \\
    \quad+ \sum_{i=1}^N \Big( Q_i(t_k)\sum_{t}\delta_i(t) + R_i^{\delta\delta} + R_i^{\delta\eta} \Big)
    \label{eq:frame-sum-expanded}
\end{align}
Next, we replace the remainder terms $R_i^{\delta\delta}$ and $R_i^{\delta\eta}$ with their respective upper bounds from \cref{eq:Rdd-bound-final} and \cref{eq:Rde-bound-final}. Since we seek an upper bound, we use $R_i^{\delta\delta} + R_i^{\delta\eta} \le |R_i^{\delta\delta}| + |R_i^{\delta\eta}|$. This allows us to collect all terms related to the second moments of arrivals and departures:
\begin{align}
    L(Q(&t_{k+1})) - L(Q(t_k)) \le \sum_{i=1}^N Q_i(t_k)\sum_{t}\delta_i(t) \nonumber \\
    &+ C \cdot F_k \sum_t\sum_{i=1}^N \big(A_i(t)^2+D_i(t)^2\big)
\end{align}
for some positive constant $C$. The second line collects the initial term from \cref{eq:frame-sum-expanded} and the bounds on the remainders into a single expression that is linear in the frame length $F_k$ and the second moments. Finally, we take the conditional expectation of both sides with respect to $Q(t_k)$. Let $\bar{\delta}_{i,k} \triangleq \mathbb{E}[\sum_t (A_i(t) - D_i(t)) | Q(t_k)]$ be the expected net change for queue $i$ in frame $k$. This yields the final, clean bound:
\begin{align}
    \mathbb{E}[L(Q&(t_{k+1})) - L(Q(t_k)) \,|\, Q(t_k)] \nonumber \\
    &\le B \cdot \mathbb{E}[F_k|Q(t_k)] + \sum_{i=1}^N Q_i(t_k)\bar{\delta}_{i,k}
    \label{eq:frame-drift-final}
\end{align}
where the constant $B$ absorbs $C$ and the expected second moments of the arrival and departure processes, which are assumed to be bounded. By collecting all second-moment disturbance terms into a single constant and expressing the net change in terms of system parameters, we arrive at the final bound:
\begin{align}
\label{eq:frame_drift_bound}
\Delta_k \le{}& B \cdot \mathbb{E}[F_k \mid Q(t_k)]  \\
&+ \sum_{i=1}^N Q_i(t_k) \left( \lambda_i \Delta t \, \mathbb{E}[F_k \mid Q(t_k)] - \mathbb{E}[\mu_i(k) \mid Q(t_k)] \right)\nonumber
\end{align}
Here, $B$ is a finite constant independent of the control policy, $\lambda_i$ is the arrival rate for queue $i$, $\Delta t$ is the time-slot duration, and $\mu_i(k) \triangleq \sum_t D_i(t)$ is the total service delivered to queue $i$ during frame $k$.

We now outline the proof that ACI is throughput-optimal. The proof depends on showing that for any arrival rate vector $\Lambda$ strictly within a scaled version of the capacity region, the Lyapunov drift is negative. 

\subsubsection{Constant-Factor Approximation Lemma}
The ACI rule from \cref{eq:aci_rule} maximizes a score that includes the switching modulator $f_{ij}(t)$. Let the unscaled, per-unit-time goodput objective be:
\begin{equation}
    \mathcal{G}(i,L \mid t_k) \triangleq \frac{Q_i(t_k) \cdot \mathbb{E}[\mu_i(k) \mid Q(t_k)]}{ \mathbb{E}[\tau_{ij}(t_k) + L \Delta t \mid Q(t_k)] }
\end{equation}
The ACI policy selects $(i^*, L^*)$ to maximize the scaled objective $\mathcal{H}(i,L) \triangleq \mathcal{G}(i,L) \cdot f_{ij}(t_k)$. Because the modulator is bounded such that $f_{ij} \in [f_{\min}, f_{\max}]$, the chosen action $(i^*, L^*)$ is guaranteed to achieve at least a constant fraction $\zeta \triangleq f_{\min}/f_{\max}$ of the maximum possible unscaled objective.
\begin{equation}
    \mathcal{G}(i^*,L^* \mid t_k) \ge \zeta \cdot \max_{i,L} \mathcal{G}(i,L \mid t_k)
    \label{eq:fractional-argmax}
\end{equation}
This holds because $\mathcal{G}(i^*,L^*) f_{ i^* j} \ge \mathcal{G}(\hat{i},\hat{L}) f_{\hat{i} j}$, where $(\hat{i},\hat{L})$ maximizes $\mathcal{G}$. Rearranging and using the bounds on $f$ proves the lemma.

\subsubsection{Frame Drift Inequality and Negative Drift}
\begin{figure}[!t]
    \centering
    \includegraphics[keepaspectratio, width=\columnwidth, height=3cm]{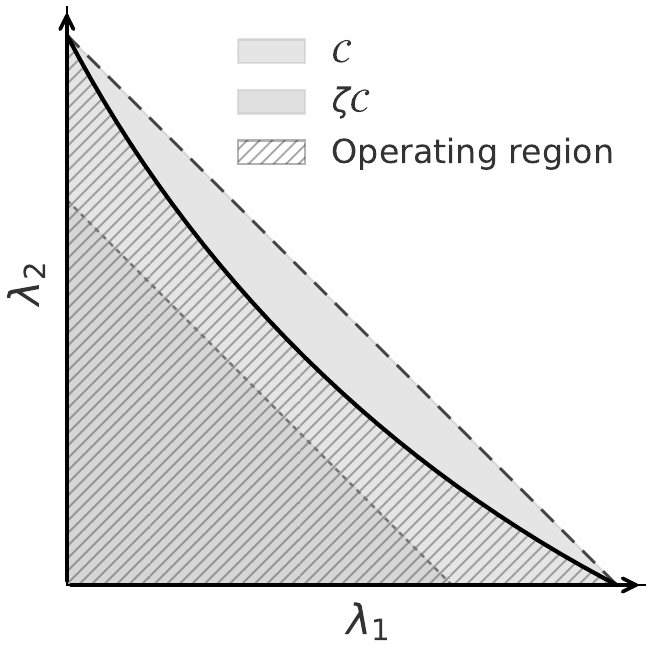}
    \caption{Capacity region: with switching ($\zeta C$) vs without ($C$)}
    \label{fig:capacity_region}
\end{figure}

We begin with the frame-drift bound from \cref{eq:frame_drift_bound}. By identifying $\mathbb{E}[\mu_{i^*}(k) \mid Q(t_k)]$ as the service provided by ACI and dividing by the expected frame duration $\mathbb{E}[\Delta t \cdot F_k \mid Q(t_k)]$, we can relate the drift to the ACI objective:
\begin{equation}
\frac{\Delta_k}{\mathbb{E}[\Delta t \cdot F_k\mid Q(t_k)]} \le B' + \sum_{i=1}^N Q_i(t_k)\lambda_i - \mathcal{G}(i^*,L^*\mid Q(t_k))
\label{eq:per-unit-time-drift}
\end{equation}
Applying the lemma from \cref{eq:fractional-argmax}, we can lower-bound the service term by what an optimal stationary policy could achieve. For any target rate vector $\bar{\boldsymbol{\mu}} \in \mathcal{C}$, there exists a policy that yields an objective value of $\sum_i Q_i(t_k)\bar{\mu}_i$. Therefore:
\begin{align}
\frac{\Delta_k}{\mathbb{E}[\Delta t \cdot F_k \mid Q(t_k)]}
&\le B' + \sum_{i} Q_i(t_k)\lambda_i - \zeta \sum_{i} Q_i(t_k)\bar{\mu}_i \nonumber \\
&= B' - \sum_{i} Q_i(t_k)\bigl(\zeta\bar{\mu}_i-\lambda_i\bigr)
\label{eq:drift-zeta-form}
\end{align}
For the drift to be negative for large queue lengths, the term in the parenthesis must be positive. This requires that the arrival rate vector $\Lambda$ lies strictly inside the $\zeta$-scaled capacity region, $\zeta\mathcal{C}$. Specifically, for some $\alpha > 0$, we need a feasible service vector $\bar{\boldsymbol{\mu}} \in \mathcal{C}$ such that:
\begin{equation}
    \zeta\bar{\mu}_i \ge \lambda_i + \alpha \quad \text{for all } i
\end{equation}
Under this condition, the drift becomes negative for sufficiently large $\sum_i Q_i(t_k)$, which guarantees network stability by the Lyapunov criterion. The negative drift condition proves that the ACI policy stabilizes any arrival rate vector $\Lambda$ within the interior of the scaled capacity region $\zeta\mathcal{C}$. This establishes the throughput optimality of the ACI framework. \Cref{fig:capacity_region} shows how switching delays shrink the ideal capacity region $\mathcal{C}$. Since ACI avoids costly switches unless necessary, its operating region for any stabilizable load $\Lambda$ lies between these two boundaries.

\subsection{Guaranteed Switching and Starvation Avoidance}
A key property of ACI is that it avoids starving any queue with persistent arrivals. We prove this by showing that if the algorithm stays on a single queue $i$ for too long, the score for switching to another queue $j$ will inevitably become dominant, forcing a switch. At a decision epoch $t_k$, let the currently served queue be $i$. The ACI algorithm compares the score of staying on queue $i$ versus switching to any other queue $j \neq i$. To analyze this trade-off, we can simplify the ACI score from \cref{eq:aci_rule} to isolate the core components.

The score for \textbf{staying} on queue $i$ (where the switching time $\tau_{ii}=0$) is proportional to its queue-weighted rate:
\begin{equation}
    \text{Score}_{\text{stay}}(i) \propto Q_i(t_k) \cdot \mathbb{E}[R_i] \cdot f_{ii}(t_k)
\end{equation}
The score for \textbf{switching} to queue $j$ for a dwell of $L$ slots must account for the initial switching delay $\tau_{ij}$, which amortizes the goodput:
\begin{equation}
    \text{Score}_{\text{switch}}(j, L) \propto Q_j(t_k) \cdot \mathbb{E}[R_j] \cdot \frac{L \cdot \Delta t}{\tau_{ij}(t_k) + L \cdot \Delta t} \cdot f_{ij}(t_k)
\end{equation}
A switch to $(j, L)$ is guaranteed if $\text{Score}_{\text{switch}}(j, L) > \text{Score}_{\text{stay}}(i)$. Rearranging this inequality shows that a switch is forced whenever the backlog $Q_i(t_k)$ exceeds a specific threshold $\Theta_{ij}(L)$:
\begin{equation}
    Q_j(t_k) > \Theta_{ij}(L) \triangleq Q_i(t_k) \cdot \frac{\mathbb{E}[R_i] f_{ii}}{\mathbb{E}[R_j] f_{ij}} \left(1 + \frac{\tau_{ij}(t_k)}{L \cdot \Delta t}\right)
    \label{eq:switching_threshold}
\end{equation}
Crucially, this threshold is finite for any queue $i$ with a non-zero expected rate ($\mathbb{E}[R_i] > 0$). The threshold is lowest for a long dwell ($L \to \infty$), but it remains finite even for the smallest possible dwell, $L=1$.

Now, consider a stabilizable system where queue $i \neq j$ has a persistent arrival rate $\lambda_i > 0$. If the scheduler were to repeatedly decide to stay on queue $i$, queue $j$ would receive no service. Consequently, its backlog $Q_i(t)$ would grow, on average, with every passing frame. By the law of large numbers, the backlog $Q_i(t)$ is guaranteed to eventually cross the finite threshold $\Theta_{ij}(L)$ for any potential dwell $L$.

At that point, the ACI rule \emph{must} select a switch to queue $j$ to maximize its score. This proves that ACI cannot indefinitely ignore a queue with incoming data, guaranteeing that no queue is starved and that switches are inevitable in a live system.

\subsection{ACI Urgency Metric Variations}
\label{subsec:aci_variations}
ACI is flexible, so the backlog term $Q_i(t)$ can be replaced by Head-of-Line (HoL) age to target latency. We consider two variants, \textbf{ACI-A (Age-Aware)} with score $\mathrm{Age}_i \cdot \bar{\mu}_{i\mid j}$ and \textbf{ACI-PA (Pure-Age)} with score $\mathrm{Age}_i$ only. These variants can reduce latency, yet they are not guaranteed to be throughput-optimal. The backlog-driven ACI is throughput-optimal because the decision weight $Q_i(t)$ aligns with the Lyapunov $L(Q)=\tfrac{1}{2}\sum_i Q_i^2$ and ensures negative drift under stabilizable load. When backlog is replaced by HoL this alignment disappears, the scheduler may serve a very old packet on a near-zero-rate channel, negative drift is no longer guaranteed, and other queues can grow without bound at high load.

\section{Evaluation}
\label{sec:evaluation}

We evaluate ACI on a simulated FSO backhaul for multi-UAV networks. Key physical and system parameters are listed in \Cref{tab:sim_params}.

\begin{table}[h!]
\centering
\caption{Key Simulation Parameters}
\label{tab:sim_params}
\begin{tabular}{ll@{\hspace{2em}}ll}
\toprule
\textbf{Symbol} & \textbf{Value} & \textbf{Symbol} & \textbf{Value} \\
\midrule
$N$ & 6 & $C_n^2(h)$ Model & HV-5/7 \\
$Z_m, Z_s$ & 500 m, 250 m & $A_0$ & $1.7\text{e-}14$ m$^{-2/3}$ \\
$R_{\text{loiter}}$ & 150 m & Wind Model & Bufton \\
$\Delta t$ & 10.41 ms & $\lambda$ & 1.55 \textmu m \\
$L$ & 3 slots & $P_t$ & 22 dBm \\
$\beta, \gamma$ & 1.0, 1.0 & $\sigma_p, \sigma_\theta$ & 0.05 m, 1 mrad \\
$\Lambda$ & 350 Mbps & $\theta_{\text{FOV}}$ & 9 mrad \\
$t_{\text{pilot}}, t_{\text{FSM}}$ & 1 ms, 3 ms & $v_\text{max}$ & 120 deg/s \\
$a_\text{max}$ & 600 deg/s$^2$ & $j_\text{max}$ & 4000 deg/s$^3$ \\
\bottomrule
\end{tabular}
\end{table}

\begin{figure}[htb]
    \centering
    \includegraphics[keepaspectratio,width=\columnwidth,height=4cm]{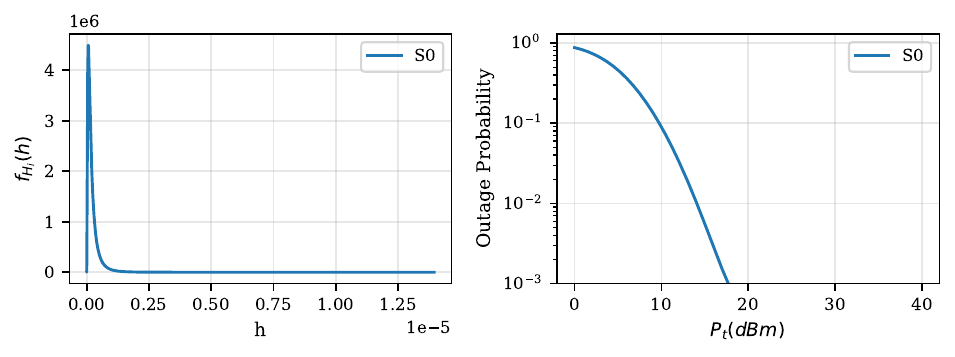}
    \caption{E2E channel PDF and outage of a single slave}
    \label{fig:channel}
\end{figure}

\begin{figure}[htb]
    \centering
    \includegraphics[keepaspectratio,width=\columnwidth,height=4cm]{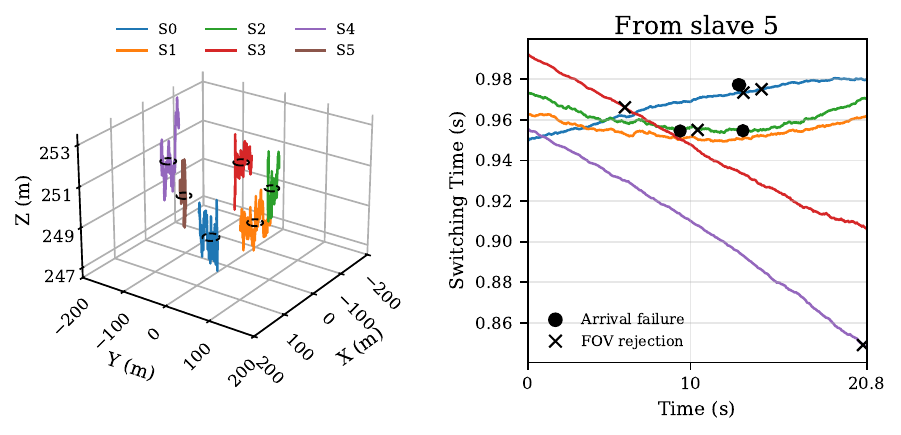}
    \caption{Slave UAV mobility and switching dynamics}
    \label{fig:switching}
\end{figure}

As shown in \Cref{fig:channel}, the E2E gain is heavily skewed toward low values, causing frequent outages at modest transmit powers. Thus, we match the slot length to the coherence time to track block fading.
The switching delays are stochastic, a consequence of UAV mobility and beam pointing errors like FOV misses and acquisition failures, as validated in \Cref{fig:switching} for a representative slave UAV.

\begin{figure}[htb]
    \centering
    \includegraphics[keepaspectratio,width=\columnwidth,height=4cm]{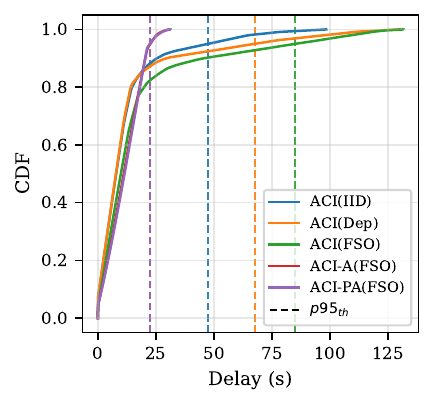}
    \caption{Overall delay CDF for ACI variants}
    \label{fig:cdf}
\end{figure}

To isolate how switching-time uncertainty shapes performance, we hold arrivals, slotting, horizon, and per-slot rates fixed and compare three switch-time models. In the \emph{IID} case, switch times fluctuate around ring-distance means (near/mid/far) with no memory—think “roughly the same cost every time.” The \emph{dependent} case introduces AR(1)-type drift at two levels (global and per-target), so the system experiences recognizable fast/slow phases—bursts of cheap or expensive switches that persist for a while. The \emph{FSO-driven} case mirrors that correlated behavior but adds the realities of optics: geometric FOV misses and probabilistic acquisition retries. Crucially, \Cref{fig:cdf} reports \emph{packet-delivery  delay}, not switching time; switching costs affect it \emph{indirectly} by interrupting service and changing when the scheduler chooses to dwell or move. Consistent with this, ACI(FSO) tracks ACI(Dependent) across most quantiles and then separates in the upper tail due to retries and FOV rejections; ACI(IID) is similar near the median and lightest-tailed because it lacks bursts and failures. The age-prioritizing variants compress the tail: ACI-A (channel-aware age) and ACI-PA (pure age) nearly overlap, with ACI-A slightly left-shifted from better tie-breaking when rates differ.

\begin{figure}[htb]
    \centering
    \includegraphics[keepaspectratio,width=\columnwidth,height=4cm]{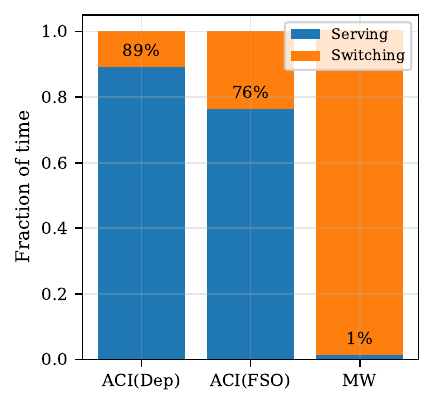}
    \caption{Fraction of time spent serving vs.\ switching}
    \label{fig:time}
\end{figure}

The time budget in \Cref{fig:time} shows the mechanism. A switching-blind Max-Weight chases instantaneous rates, retargets often, and pays a full slew–pilot–settle cycle each time, so wall-clock time is spent switching and service collapses to about 1\%. ACI prices switches, rewards adjacency, and commits in short frames, banking dwell until the expected gain exceeds the penalty. As a result, ACI(Dependent) converts about 90\% of time into service, while ACI(FSO) remains service-dominant but lower at 75–80\% because acquisition retries and FOV misses add overhead. The Dependent–FSO gap is the tax of stochastic acquisition failures beyond pure kinematics.

\begin{figure}[htb]
    \centering
    \includegraphics[keepaspectratio,width=\columnwidth,height=4cm]{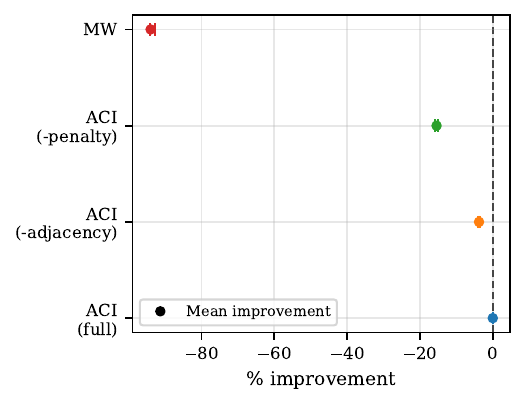}
    \caption{Effect of ACI components on mean delay (vs. MW)}
    \label{fig:ablation}
\end{figure}

To understand each ACI element, we use mean delay as the improvement metric. \Cref{fig:ablation} shows that removing adjacency (longer slews) or the switch penalty (flapping from repeated slew–pilot–settle cycles) raises delay, while the full policy attains the lowest mean delay—evidence that both proximity cues and explicit switch-time pricing are necessary.

\section{Summary}
We address the problem of dynamic server allocation to parallel queues with stochastic, inhomogeneous switch costs and time-varying connectivity, a problem motivated by FSO backhaul in multi-UAV networks. To overcome the inefficiency of the myopic Max-Weight policy, we propose ACI, a non-myopic, frame-based framework to amortize these switching delays.
We prove the backlog-driven ACI is throughput-optimal for a scaled capacity region. Further analysis shows that age-based variants provide superior latency, which reveals a clear performance trade-off: the standard ACI is ideal for maximum throughput, while the channel-aware ACI-A offers the best balance for latency-sensitive applications.

\bibliographystyle{IEEEtran}
\bibliography{bibliography}
\end{document}